# Amino Acid Distributions and the Effect of Optimal Growth Temperature

Benjamin D. Greenbaum[1], Pradeep Kumar[2], and Albert Libchaber[1,2]

[1]*The Simons Center for Systems Biology, Institute for Advanced Study, Princeton, New Jersey, USA*

[2]*Center for Studies in Physics and Biology, The Rockefeller University, New York, New York, USA*

Corresponding Author: Benjamin Greenbaum, beng@ias.edu

**Abstract**

We perform an exhaustive analysis of genome statistics for organisms, particularly extremophiles, growing in a wide range of physicochemical conditions. Specifically, we demonstrate how the correlation between the frequency of amino acids and their molecular weight, preserved on average, typically decreases as optimal growth temperature increases. We show how the relation between codon degeneracy and amino acid mass is enforced across these organisms. We assess the occurrence of contiguous amino acids, finding several significant short words, often containing cysteine, histidine or proline. Typically, the significance of these words is independent of growth temperature. In a novel approach, first-passage distributions are used to capture correlations between discontiguous residues. We find a nearly universal exponential background that we relate to properties of the aforementioned individual amino acid frequencies. We find this approach reliably extracts correlations that depend on growth temperature, some of which have not been previously characterized.

**Introduction**

To what extent is genomic amino acid composition broadly dictated by physical and chemical properties instead of natural selection or chance? Most studies focus on the effect of amino acid alterations on functional properties of proteins or gene families. Yet the landscape upon which an organism evolves is ultimately related to constraints on its amino acid repertoire. Likewise, the existence of universal patterns in amino acid usage across vast evolutionary distances could indicate fundamental properties relevant to all life.

The genome sequencing of many extremophiles has created an opportunity to probe these basic, yet practical questions[1]. A large diversity of microbes thrives in extremes of physicochemical conditions such as pressure, temperature, and pH[2]. Extremophiles have been sequenced at both vast evolutionary distances and across a broad range of environmental conditions. Hence these organisms provide a deep-field lens for resolving how variations in physiochemical environment alter genome characteristics. One common feature of these extreme conditions is their tendency to denature enzymes from organisms living at normal conditions, yet proteins from extremophiles must be stable and functional. Stability of proteins relies on various features of amino acid sequences, and the adaptive mechanisms to thrive on these conditions are presumably many and varied. Motivated by this, several authors have noted various effects of optimal growth temperature (OGT) on certain amino acid features[3,4,5,6,7].

In this work, we present a general, hierarchical approach to the issue by exhaustively studying statistical properties of the proteome of 76 organisms. These organisms, largely extremophilic, were chosen for the wide range of environmental conditions they inhabit, along with a set of mesophilic genomes selected for comparison. We investigate the properties of individual amino acids and their underlying nucleotides, and use these properties to assess the significance of contiguous amino acid words and correlations between discontiguous residues. We find that while a linear relationship between amino acid frequency and mass holds on average across these organisms[8], the strength of correlation weakens at high OGT. The average distribution is reminiscent of the early Miller experiments[9], where 95% of the amino acids produced were the four lightest, leading Eigen to propose an origin of life with these four amino acids[10]. Cysteine and leucine are the only two exceptions to this rule, with cysteine on average less frequent than one would expect given its mass and leucine more.

The degeneracy in the genetic code is itself related to amino acid mass - smaller degeneracy implying heavier amino acids[11]. While the heavier amino acid arginine is an apparent exception, having a six-fold degenerate coding, we show that its codon usage is effectively only three- to four-fold across these organisms. The two classes of aminoacyl-tRNA synthetase utilized by each amino acid are somewhat related to mass. Class 2, the oldest one is associated with the lighter amino acids. The lightest amino acids are on average more degenerate and may be associated with the oldest synthetase.

We show how these observations carry over to properties of amino acid words[12] and first-passage distributions for correlations between discontiguous residues. While the former shows mainly OGT independent relative abundance measures, significant values for the latter are strongly OGT dependent. We observe a universal exponential background distribution for first-passage times that we relate to our initial observations concerning individual amino acid frequencies and OGT. By understanding this background distribution we are able to reliably extract a set of non-random correlations, all OGT dependent. Some likely relate to known thermal adaptations, such as disulfide bonding and salt-bridge formation. Yet our approach also captures unexplained, OGT-dependent effects. Hence careful exploration of these issues may yield unforeseen phenomena from extremophile genomes.

## Results

### Individual Amino Acids Statistics and their Physical and Evolutionary Implication

We examine the distribution of individual amino acids, $\{a_i\}_{i=1}^{20}$, across the entire proteomes of 76 organisms (Supplementary Table 1)[13,14]. In addition to several model organisms, such as humans, mice, and yeast, these organisms were chosen based on the broad range of environmental conditions they inhabit. In Fig. 1a, we show the observed frequency of all 20 amino acids, $f(a_i)$, versus molecular weight, for these organisms, highlighting the median values. Molecular weight correlates negatively with observed frequency, with the strong exceptions of cysteine and leucine. Excluding these two amino acids the linear correlation coefficient between median frequency and molecular weight, $r_{fW}$, is -0.6814. However, the strength of correlation is less than what one would obtain using standard tables for mesophiles[8,15]. Using one such table, we find a correlation of -0.7899 for these same 18 amino acids[8]. Hence our observation suggests a bias in the frequency of amino acids that may also lead to amino acid correlations, a theme we will explore throughout this analysis.

The change in frequency is also reflected in the genetic code degeneracy for each amino acid, as shown in Fig. 1b[11,16]. Apart from arginine, observed frequency sorts the amino acids into those with high degeneracy and low mass, and those with low degeneracy and heavier mass. Using an entropic measure of codon usage, we show that arginine strongly violates this trend - its degeneracy is essentially between 3 and 4 codons (Supplemental Methods). This is consistent with codon-based arguments for the relative newness of arginine in the genetic code[17]. The change in frequency is also apparent in the class of aminoacyl-tRNA synthetase utilized by each amino acid[18]. Fig. 1c presents the synthetase class used (type 1 and type 2). Both ends of the mass spectrum correspond to the class of aminoacyl-tRNA synthetase used, with the ancestral class 2 synthetase dominating the light, more degenerate end, and class 1 the heavier, less degenerate end.

To understand better how amino acid frequencies vary, we look to an organism's OGT. Its value was taken from the standard extremophile reference and references therein[1]. If a OGT range was given we used the midpoint. In Fig. 2a, we plot the relationship between $r_{fW}$ and OGT for all amino acids other than cysteine and leucine. As a general rule, the correlation between molecular weight and observed frequency decreases with OGT. As a demonstration of this effect, in Fig. 2b we plot the relationship between molecular weight and amino acid frequency for the same 18 amino acids for one low OGT species, *P. arcticus* ($r_{fW}$=-0.8016), and one high OGT species, *S. azorense* ($r_{fW}$=-0.3841). For the lower temperature organism the correlation is strong while for the high temperature organism it is weakened.

Others have focused specifically on the relationship between specific amino acid frequencies and OGT[3,4,5]. In Supplementary Table 2, we show the linear correlation coefficient between amino acid frequency in each organism and OGT, $r_{fT}$. A group of class 1 amino acids, particularly IVEY, are positively correlated, while QTHD are similarly negatively correlated. Save D, these were noted in a smaller study[4]. The negative correlation with OGT of many class 2 amino acids was separately noted, though we find the strongest negative correlate, Q, is class 1[5]. The analysis of Zeldovich, *et al.*, who associated the increased in class 1 amino acid frequency in thermophiles with protein stability, may therefore drive these observations[3].

Given that global amino acid frequency is strongly linked to molecular weight, we examined the relationship between the frequencies of amino acids and the frequencies of the nucleotides that encode them. We define $\xi(a_i)$ as

$$\xi(a_i) = \log\left(\frac{f(a_i)}{\sum_{j=1}^{d_i} f(1_j)f(2_j)f(3_j)}\right)$$

where $d_i$ is the degeneracy of amino acid $a_i$ and $f(1_j)$ is the frequency across the entire coding region of the nucleotide encoding the first position of the *j*-th codon, hence the denominator is the expected value of $f(a_i)$, assuming random placement of coding nucleotides. The value of $\xi(a_i)$ for all 20 amino acids across the 76 aforementioned genomes is shown in Supplementary Fig. 1. The expected value is often a surprisingly close approximation, save for cysteine. Our results largely support Zeldovich, *et al.,* who conclude that nucleotide content in coding regions is largely driven by amino acid usage[3].

**Relative Abundance of Amino Acid Words**

Once the occurrence of amino acid frequencies is established across this broad range of organisms, we examine how well these frequencies explain the occurrence of amino acid words. We look at the distribution of the logarithm of relative abundances for dimers, $\omega_{ij}$, according to the formula

$$\omega_{ij} = \log\left(\frac{f(a_i a_j)}{f(a_i)f(a_j)}\right)$$

and trimers, $\omega_{ijk}$, according to

$$\omega_{ijk} = \log\left(\frac{f(a_i a_j a_k)f(a_i)f(a_j)f(a_k)}{f(a_i a_j)f(a_j a_k)f(a_{i\_}a_k)}\right)$$

We assess significance by treating $\omega_{ij}$ as normally distributed[12]. In Table 1, we list the amino acid pairs whose median Z-score for all organisms is greater than two (where a Z-Score for one organism is the number of standard deviations from the mean value for all dimers in that organism), along with the number of organisms for which this occurs. The CC pair has the highest median, and is more than two standard deviations away from its mean in 61 of the genomes we observed. Cysteine also is involved in 3 of the 6 overrepresented pairs shown (CC, CP and CH). The other three pairs are HH, HP, and WW. Cysteine and histidine rich domains are both associated with metal-ion binding[19,20], and both are favored to appear next to themselves, next to each other and next to proline. The only pair consistently under-represented at the same strength is GP, which represents the pairing of an amino acid associated with rigidity (P) with one associated with flexibility (G)[21]. They are the two amino acids that do not follow the typical Ramachandran plot[22]. The general over-representation of dimers involving C, H and W in a smaller set of genomes was noted by Karlin[12].

Supplementary Table 4 shows an equivalent table for $\omega_{ijk}$. Several trimers are not present in any genome, indicating the rapidly waning statistical power of this method at even short lengths. As a consequence, when a rare trimer does appear it may have a wide variance, so the table indicates deviations of greater than four. Many of the most over-represented trimers again contain C and H residues.

Even though our approach of looking across many genomes looses power quickly, for organisms with larger coding regions one gains statistical power. Fig. 3 shows the distribution of these values for *A. thaliana*, which has the longest genome and many of the longest genes of any organism studied here, illustrating the power of these methods when one has sufficient statistics. In Fig. 3a, $\omega_{ij}$ is plotted for all of the dimers in *A. thaliana*. The x-axis indicates an index for each dimer, $(i, j)$, where $i$ and $j$ range from 1 to 20 to index each amino acid. A large group of dimers are clearly randomly (in fact normally) distributed, with outliers immediately apparent. The largest value belongs to HH, while the lowest here is EP, with GP a close second. For trimers, the separation of signal from background is even more apparent (Fig. 3b). Here the largest value is HHH and the lowest is CHH.

We examined whether any of these values, many of which involve residues involved in structural stability, are OGT dependent. While most significant motifs in Table 1 do not display any strong OGT correlation, CP displays a very strong linear correlation (0.8088), as shown in Supplementary Fig. 2. This fact has not been noticed. We speculate that this is due to structural issues relating to cysteine usage, where a nearest-neighbor proline may be functionally relevant[19,23]. However, by and large these statistics show OGT independent effects.

**First-Passage Distributions**

To examine correlations that reflect an organism's environment, we introduce first-passage distributions to this problem: the probability that the first occurrence of amino acid $Y$ occurs $n$ residues away from amino acid $X$ without having occurred beforehand. This is denoted by

$$f_{XY}^{(n)} = p(a_n = Y; a_k \neq Y, 0 < k < n | a_0 = X),$$

where $a_i$ denotes the residue a distance $i$ from $a_0$[24].

For a given chain one can calculate the typical first-passage distribution in either the N-terminus or C-terminus direction, which we respectively denote by $f_{XY}^{(n-)}$ and $f_{XY}^{(n+)}$. If amino acids were uncorrelated, this would reduce to a geometric distribution, the discrete analog of an exponential distribution[25]. In this case the probability would only depend on the frequency of $Y$ and should, therefore, reflect the OGT dependent properties of individual amino acids, so that the OGT dependence of outliers can be more clearly established. However, amino acid probabilities are not all independent, as demonstrated in the previous section. As shown in the Supplementary Methods, the log of our distribution often becomes linear after a few residues of separation, with a slope dependent on $Y$ and not $X$. We therefore assume $\lambda_{XY}^{\pm}\exp(-\lambda_{XY}^{\pm}n)$ as our theoretical background distribution for $f_{XY}^{(n\pm)}$, where $\lambda_{XY}^{\pm}$ is derived from the best linear fit to the logarithm of the first-passage distribution versus $n$. Practically, we derive it over values of $n$ less than 50.

Previous authors have explored different measures of pairwise correlation between residues in proteomes[26,27]. Liang, *et al.* calculated the frequency of chains of length $n$ beginning with amino acid $X$ that terminate with amino acid $Y$, divided by the frequency of $Y$ in the proteome. Rosato, *et al.* calculated a similarly motivated odds ratio by comparing the number of times amino acid $Y$ occurs $n$ residues away from amino acid $X$ and dividing this by the expected number of times this would occur in a random proteome with the same length, number of proteins, and amino acid frequencies[28]. Our method does not assume individual amino acid independence and separates true discontiguous correlations from those due to many small words by excluding events where $Y$ previously occurred. It derives significance from the exponentiality of these distributions ``in the long run'', accounting for the likely dependence of the background on OGT. As shown in the Supplementary Methods, $\lambda_{XY}^{\pm}$ typically depends on $Y$ and varies with OGT when the frequency of $Y$ depends on OGT. If $Y$ is more abundant at high OGT amino acids will typically pass to $Y$ faster, and if it is less abundant more slowly. $f_{CC}^{(n\pm)}$ is a strong exception. Though less frequent at high OGT, cysteines average return time is faster - indicating that as cysteines become rare having them nearby at high OGT is more critical.

As a result of our improved background, all of our significant deviations from the expected exponential distribution depend on OGT. We extract these events from the logarithm of the ratio of the empirical value of $f_{XY}^{(n\pm)}$ to its expected value from the exponential distribution, termed $\phi_{XY}^{(n\pm)}$. Table 2a looks at the most commonly over-represented pairs, while Table 2b looks at the same information for under-represented pairs. Those shown are at least three standard deviations from the mean for a minimum of 25 of the organisms studied. $\phi_{CC}^{(3-)}$ is the most significant, by a substantial margin, both in magnitude and number of organisms in which it is over-represented. Several over-represented pairs fall into categories, such as having a positively and negatively charged side chain (RE, KE, ER, EK). Leucine, the amino acid whose frequency is higher than expected from its mass, is frequently in over-represented pairs (LE, LQ, LK, LR).

We examine in detail a set of correlations along the N-Terminal direction, all of which have C-Terminal analogs. We focus on four cases from Table 2: $\phi_{CC}^{(3-)}$, $\phi_{RE}^{(3-)}$ (which has similar behavior to $\phi_{KE}^{(3-)}$), $\phi_{LE}^{(2-)}$ (the most commonly over-represented Leucine containing pair), and $\phi_{LK}^{(2-)}$, a correlation previously unobserved to our knowledge. These pairs have the highest values of $\phi_{XY}^{(n\pm)}$ and have distinct behavior from each other.

Fig. 4 shows the number of first-passages as a function of $n$ for CC and RE for two organisms, humans and the thermophile *T. petrophila*. In both cases, $N_{CC}^{(3-)}$ and $N_{RE}^{(3-)}$ stand out from the background distribution. In the human genome, which is both longer and has longer genes, additional structures at larger $n$ appear. However, the relative height of the peak to the background distribution is higher for the thermophile, a pattern we now explore further.

**Dependence of Quantities on Habitat**

To determine which signals vary with OGT, we plot the four aforementioned cases as a function OGT for all organisms and calculate the linear correlation, $r_{T\phi}$. Fig. 5 shows the dependence between $\phi_{CC}^{(3-)}$ and OGT ($r_{T\phi}$=0.7054) along with $\phi_{RE}^{(3-)}$ ($r_{T\phi}$=0.7724) and $\phi_{LE}^{(2-)}$ ($r_{T\phi}$=0.8013). For $\phi_{CC}^{(3-)}$, OGT dependence holds for all non-eukaryotes. Eukaryotes are in the far left cluster. Without them the linear correlation is stronger. Giuliano, *et al.* noted a similar effect with weaker correlation[28].

The significance of $\phi_{CC}^{(3-)}$ has clear interpretations. Increased clustering of cysteines may lead to an increase in disulfide bonds that can be used more frequently per residue for thermal stability or correlated effects from metal-ion binding are more prevalent[29,30,31,32,33,34]. For instance, many high-temperature environments tend to be more acidic. Since low pH inhibits thiol-disulfide exchange, one may expect that cysteine clustering would increase in the proteome of an organism in an acidic environment to increase stability. In any case, given the rarity of cysteines at high OGT, the size of this effect is clearly of great importance for thermophiles.

Less noted, is the OGT dependence of the $\phi_{RE}^{(3-)}$. The two are oppositely charged, and may represent the stabilization of proteins via salt-bridge formation[35,36,37,38]. Intriguingly, we find that a great deal of information can be drawn from the unusual metabolism of organisms farthest from the best-fit line between $\phi_{RE}^{(3-)}$ and OGT. Those furthest from that line in the upper half-plane are *S. azorense* and *I. aggregans*, and in the lower half-plane are *G. metallireducens* and *G. sulfurreducens*. This may suggest a method by which unusual metabolic properties could be uncovered by deviation from a correlation line. Both the interpretation of why that particular amino acid correlation is proportional to OGT, and why sulfur respiration seems reflected in outliers, contain biological insight. While the former likely relates to an organism's utilization of salt-bridge formation, the later suggests a statistical measure for prediction of novel metabolism, which has practical uses such as in microbial cleaning of contaminated soil[39,40].

For the other two pairs, previously unnoted, we did not find a clear explanation for their significance. $\phi_{LE}^{(2-)}$ has the strongest linear correlation with OGT. Surprisingly, the plot of $\phi_{LK}^{(2-)}$ versus OGT (Fig. 5d), which has weaker temperature dependence ($r_{T\phi}$=0.6278), clusters these organisms into two main groups. One set, colored red, contains most thermophiles while the other, colored blue, contains lower OGT organisms. This suggests a statistical property can classify thermophiles from non-thermophiles. Moreover, the quantity $\phi_{LK}^{(2-)}$ lacks an obvious interpretation, so this discriminator of thermophiles from non-thermophiles may have unknown physical implications.

**Discussion**

We examine properties of amino acid distributions that would be missed if one focused on one gene, genome, organism, or physiochemical environment. For individual frequencies, excluding cysteine and leucine in the first case, a general dependence on mass and overall nucleotide usage in coding regions is shown across these organisms. Mass dependence is also reflected in the codon usage degeneracy in the genetic code, with lower degeneracy for heavier amino acid mass and, to a lesser extent, indicates the class of aminoacyl-tRNA synthetase usage. However, the strength of the known correlation of amino acid frequency with mass decreases with OGT.

The implication is that mass and OGT are the main determinants in genome wide individual amino acid usage and coding degeneracy, which then determine the overall usage of coding nucleotides. One temperature dependent, physical mechanism that relates the masses of amino acid and the origin of the genetic code was explored in Lehmann, *et al.*[41]. The direct relationship between mass and codon usage degeneracy would seem to imply an origin of the genetic code driven by amino acid chemistry in a particular environment[10], and has been used to support neutral evolutionary theories[8,11]. This viewpoint stands in some opposition to information-theoretic approaches to the code's origin[42]. However the weakening of this dependence with OGT may imply that some origin theories are more valid in a particular temperature regime. Certainly, a theory using the high frequency of light amino acids in genomes as a starting point[10], would not fare as well at high OGT where many of the heavier amino acids become more likely.

To explore larger patterns in amino acid usage we employed relative abundance measures for amino acid words. These statistics uncovered several significant words of length 2 and 3 residues across many diverse genomes. Many words involved cysteine and histidine residues, along with structural residues such as glycine and proline. The latter are disfavored to appear next to each other, while P often appears next to C and H. While this approach is limited to short words it seems to be particularly good at finding correlations that do not strongly depend on OGT.

The exponential decay of first-passage times has additional implications for the overall randomness of proteins. Surprisingly often, as indicated in Supplementary Table 5, the exponential first-passage distribution is what one would expect if an amino acid was not coupled to its neighbors. Even when this is not the case, exponentiality often holds. We believe this is analogous to the rapid mixing that can generate exponential first-passage distributions in a Markov Chain[43,44]. When an amino acid is not independent, it could appear in small, weak local groupings, enhancing the probability of staying near a small set of states before undergoing a transition to a forbidden state, making the decay time longer when independence is violated. Moreover, in microbes, noise due to small genome size may mask richer behavior such as short-range deviations from exponentiality.

First-passage distributions identified multiple OGT dependent statistical measures. Deviants from the general pattern of exponential decay indicate significant correlations that appear to be related to physical interactions such as disulfide bonding, salt-bridge formation, and sulfur metabolism. Moreover they identified a set of OGT dependent measures lacking a clear physical explanation, implying that this approach may be used to identify temperature sensitive amino acid correlations to uncover important properties of proteins in extreme environments. Further sequencing may uncover new, unexpected effects, while furthering our understanding of what is a typical protein at a given temperature.

## Materials and Methods

### Effective Codon Degeneracry

We define the effective degeneracy of an amino acid as $2^{H(a_i)}$, where $f(c_{ij})$ is the frequency of the $j$-th codon for amino acid, $a_i$, with codon degeneracy $d_i$ and

$$H(a_i) = -\sum_{j=1}^{d_i} f(c_{ij}) \log_2(f(c_{ij}))$$

is the Shannon entropy of the codon usage distribution for that amino acid[45]. The effective degeneracy is, essentially, the effective number of codons utilized for an amino acid in the genome. Its median value across all organisms studied is indicated in Supplementary Table 2, along with the amino acids molecular weights, and degeneracies. The effective degeneracy of arginine has the greatest variance and is much lower than 6, it essentially uses only between three and four codons.

### Empirical Evidence for Exponential Distribution

To test the hypothesis that the appropriate background distribution for the amino-acid first passage times defined in the text is indeed exponential, we performed the following set of tests. For each first-passage time distribution, $f_{XY}^{(n\pm)}$, we calculated the linear correlation coefficient between the value of $N$ and the logarithm of the first-passage time probability. A p-value was assigned to each correlation to determine the significance of that correlation using Student's t-distribution for a transformation for the correlation. Both were performed using the function ``corrcoef'' in Matlab.

In Supplementary Table 5 the $f_{XY}^{(n\pm)}$ p-values are explored for each of the 400 amino acid pairs. Even if multiple hypothesis testing is strictly taken into account, by treating each of the 76 cases of 400 hypothesis separately for a p-value cutoff of 0.05/30400 for a -log(p-value) of 13.318, there are only a 11 cases which are insignificant. All of them involve low frequency amino acids. The minimum, maximum and medium values of -log(p-value) are shown in Supplementary Table 5 for the 400 possible values of $XY$. Of the significant values, eight of them are between tryptophan and cysteine, two are between cysteine and tryptophan, and one is between methionine and cysteine. All occur in extremophiles where cysteines are rare and have large fluctuations.

Also shown in this table are the estimates for the average first-passage time, $T_{XY}^{\pm} = \frac{1}{\lambda_{XY}^{\pm}}$, as derived from the background distribution. Also listed are the standard deviation, minimum and maximum values. Specifically we show the case $T_{XY}^{-}$. In most cases the values are at least tens of residues (longer than the typical length at which Karlin statistics are significant), and depend on $Y$ only. To establish the later, we compared the median value of $T_{XY}^{\pm}$ to the values of $Y$ using the Kruskal-Wallis test, and also compare them to the values of $X$. When compared to $Y$ there was a very significant p-value of less than $10^{-71}$. When compared to $X$ there was no significance. This would imply that an analogous process the ``mixing'' mechanism for exponentiality in Markov processes is responsible for this effect[43,44]. In this case the distribution becomes stationary well before all states are explored.

If all amino acids were independently distributed according to their empirical frequencies, the first passage distribution would be a geometric distribution determined by the frequency of amino acid $Y$. In that case the $X$ independent decay constant, $\lambda_Y^G$, would be
$\lambda_Y^G = -\log(1 - p(Y))$. Associated with this is a geometric decay time, $T_Y^G$. If $T_{XY}^{\pm} > T_Y^G$, the observed exponential distribution decays faster than in the case when amino acids are independently distributed and, likewise, it decays more slowly if $T_{XY}^{\pm} < T_Y^G$. In Supplementary Table 5, we show the median value of the ratio $T_{XY}^{\pm}/T_Y^G$. In many cases the ratio is quite close to 1, indicating that the amino acids are basically independent of each other. There are several cases where the decay is much faster than in the independent case, and also a few where it is somewhat slower. However for those quantities to be calculated correctly, occurrences due to significant peaks would have to be removed, and a new background frequency inserted into the above formula.

**Temperature Dependence of Exponential Decay**

The linear correlation between the average expected time of first passage $T_{XY}^{-}$ and OGT is shown in the final column of Supplementary Table 5. The correlations largely follow the results of Supplementary Table 3, as one would expect given the exponential nature of these first passage distributions. All distributions where $Y$ is equal to V, E or Y have a negative correlation less than -0.5. These amino acids are more abundant at high OGT, and therefore an amino acid will typically arrive at them more quickly when this occurs, causing the mean first passage time to decrease. Likewise, H, T and Q, which become rare at higher OGT, all have longer mean first passage times. The strongest exception to this rule is $f_{CC}^{(n)}$. Despite the fact that cysteines typically become rare at higher OGT, its mean first-passage time becomes faster. The implication is that at high OGT, while C becomes less frequent it is more and more important that cysteines appear near each other. In addition to the significant spike at $n=3$, cysteines likely have many weak correlations with each other at higher OGT at different length scales. Such an effect only reinforces taking into account the background distribution in a manner that does not assume complete independence of residues.

**Acknowledgements**

BDG is the Eric and Wendy Schmidt member at the Institute for Advanced Study and would like to thank them for their support. He would like to thank Arnold Levine and Gérard Ben Arous for helpful discussions.

# Figures

**Figure 1: Relationship between Amino Acid Frequency and Molecular Weight**

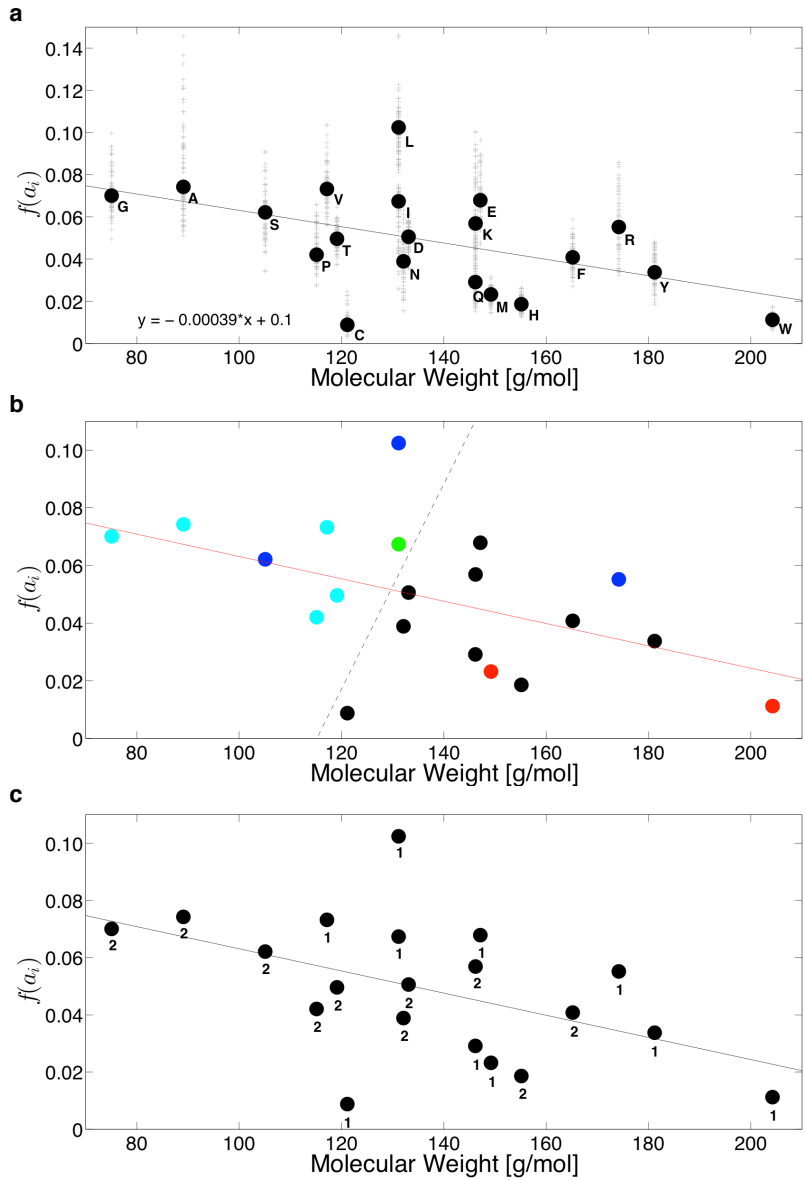

**(a)** Dependence of amino acid frequency on molecular weight for all organisms studied. The solid circles denote the median values, which can be fit with a line of slope -0.00039. **(b)** The codon degeneracy is indicated by color: light blue is degeneracy 4, blue is 6, green is 3, black is 2 and red is 1. A line separates amino acids with degeneracy greater than 4, aside from arginine. **(c)** The class of aminoacyl-tRNA synthetase utilized is labeled against the frequencies.

**Figure 2: OGT Dependence of Correlation Between Amino Acid Frequency and Molecular Weight**

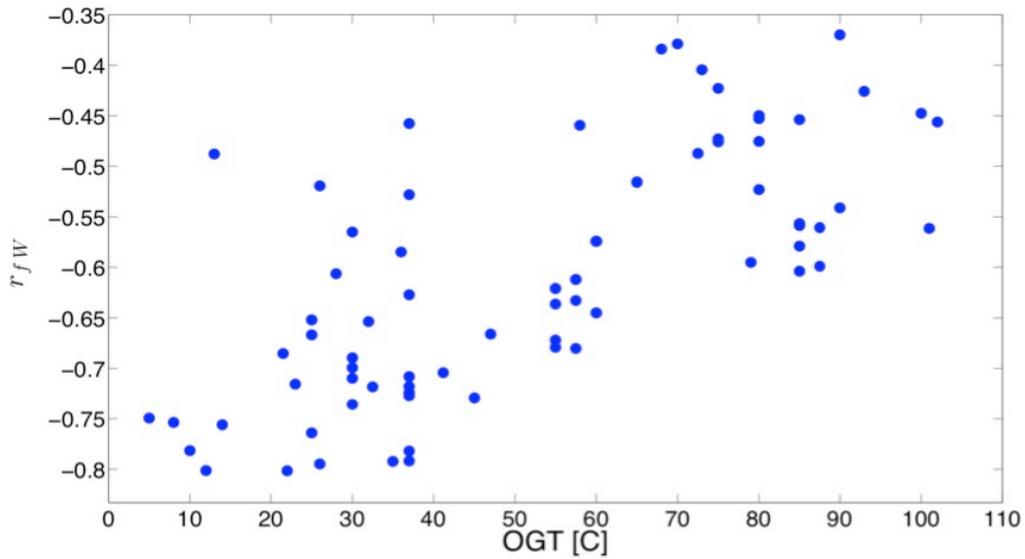

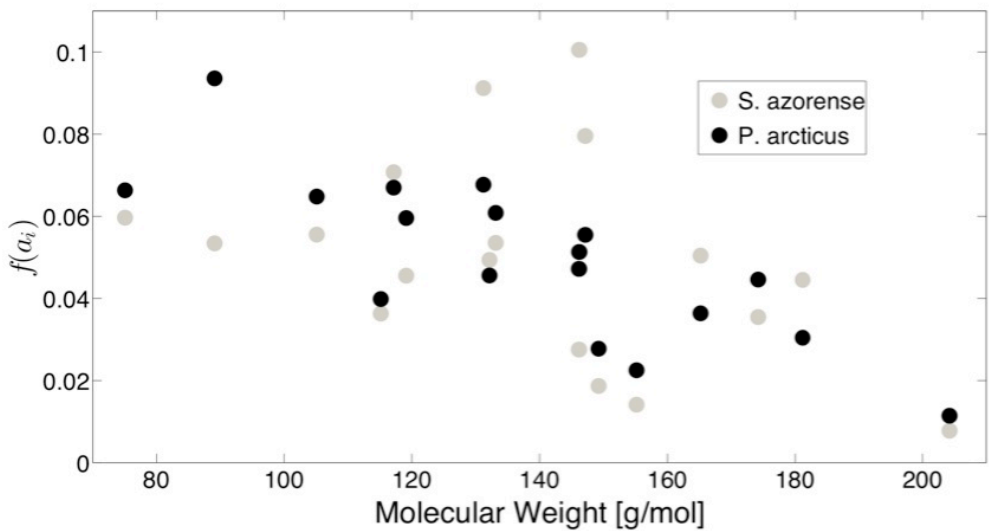

**(a)** Linear correlation coefficient between molecular weight and frequency and its dependence on OGT. **(b)** Frequency of amino acids other than cysteine and leucine as a function of molecular weight for two different organisms *S. azorense*, a thermophile, and *P. arcticus*, a psychrophile.

## Figure 3: Amino Acid Word Statistics for *A. thaliana*

**a**

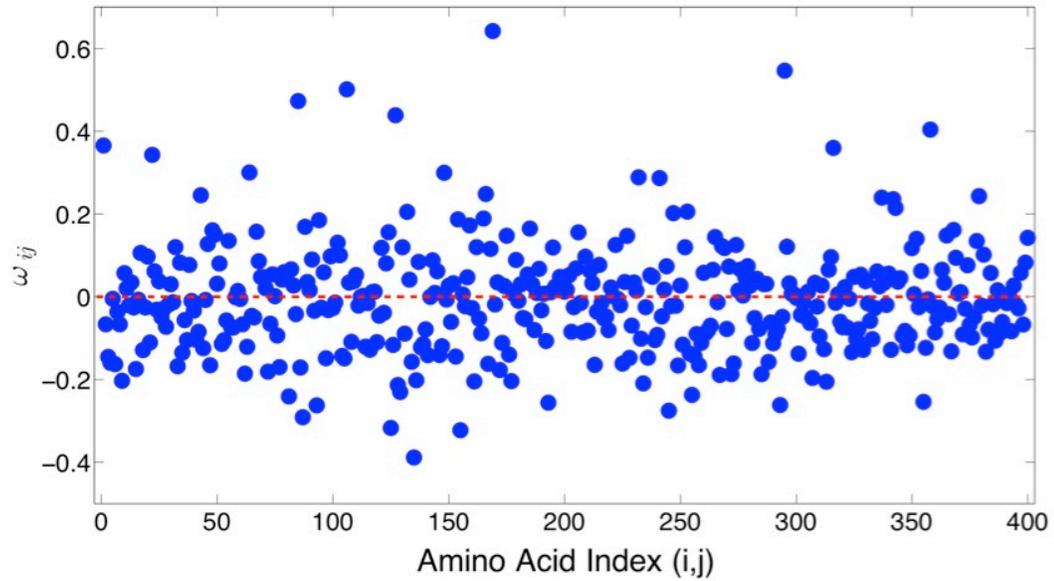

**b**

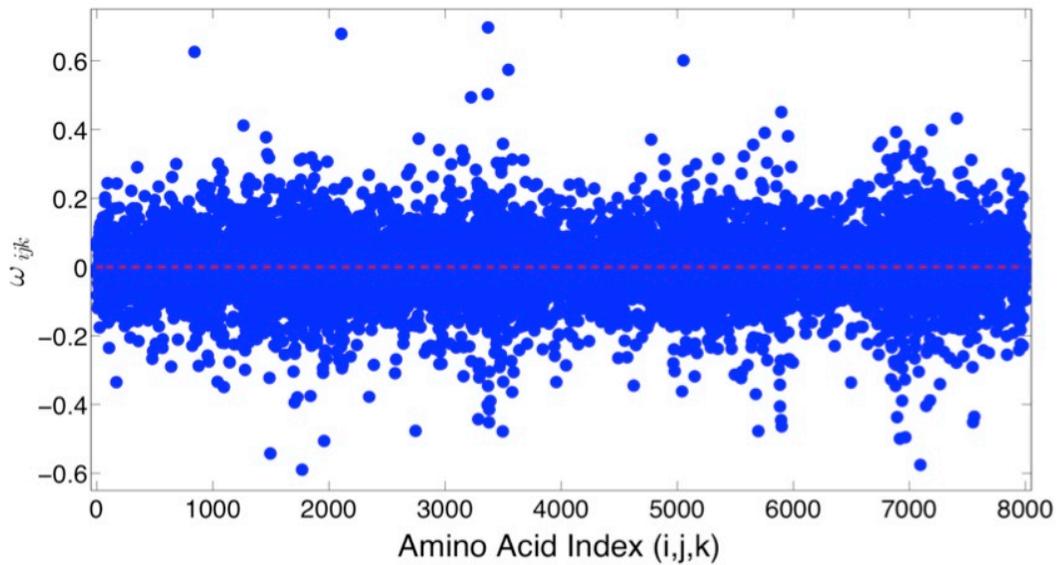

Logarithm of **(a)** dimer and **(b)** trimer relative abundance for amino acids in *A. thaliana*. Solid circles are the empirical values and the dotted reference line would indicate complete agreement between observed and expected values.

**Figure 4: Variation in Number of First-Passages from C to C and R to E**

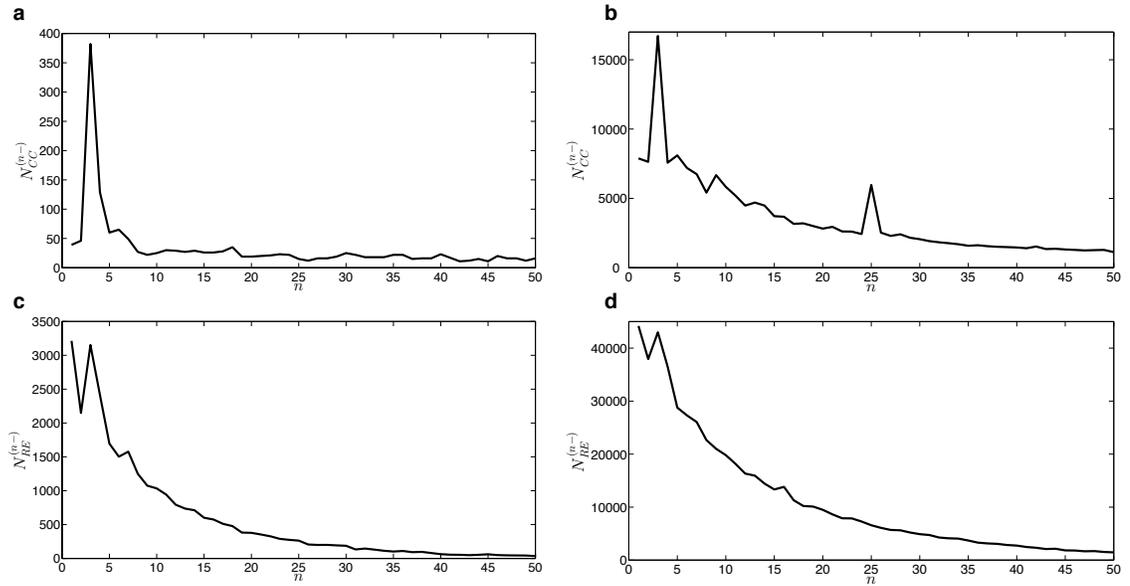

Number of first-passages from cysteine to cysteine in the **(a)** *T. petrophila* and **(b)** human proteomes, and number of first-passages from arginine to glutamic acid in the **(c)** *T. petrophila* and **(d)** human proteomes.

**Figure 5: OGT Dependence of Significant First-Passage Times**

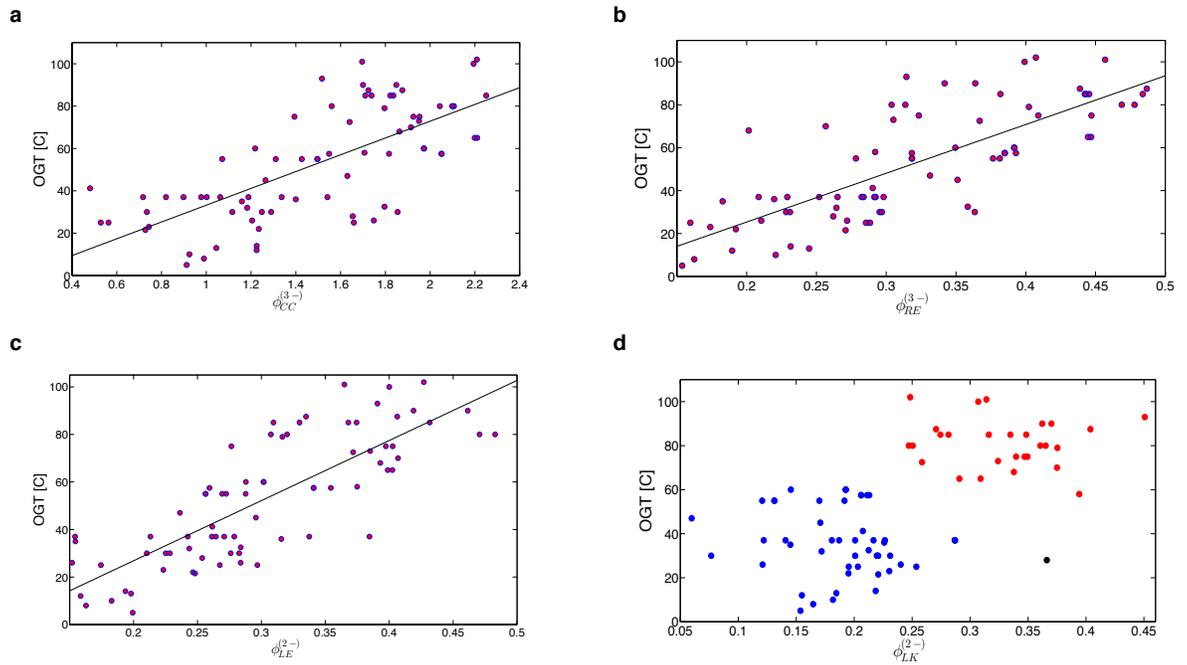

OGT as a function of the logarithm of the real to expected N-terminus first-passage probability for **(a)** CC at three residues, **(b)** RE at three residues, **(c)** LE at two residues, and **(d)** LK at two residues, where blue circles indicate OGT ranging from 5° C to about 60° C, and red circles represent hyperthermophilic organisms with OGT above 60° C.

**Tables**

**Table 1: Significant Amino Acid Pairs**

| Pair | Occurrences | Median Z-Score |
|---|---|---|
| CC | 61 | 2.8745 |
| HH | 50 | 2.2927 |
| CP | 42 | 2.2467 |
| WW | 40 | 2.2055 |
| CH | 40 | 2.1139 |
| HP | 39 | 2.0398 |
| GP | 47 | -2.18 |

## Table 2: Significant First-Passage Times

### Table 2a: Over-represented First-Passage Times

| Significant First-Passage Times | | | |
|---|---|---|---|
| N-Terminal Direction | | | |
| Pairs | Number of Genomes | n | $\phi_{XY}^{(n-)}$ |
| CC | 75 | 3 | 1.545 |
| LE | 57 | 2 | 0.2878 |
| LQ | 48 | 2 | 0.1836 |
| LK | 47 | 2 | 0.2207 |
| RE | 41 | 3 | 0.3094 |
| LR | 35 | 2 | 0.1812 |
| IQ | 34 | 2 | 0.1583 |
| KE | 32 | 3 | 0.325 |
| KM | 31 | 2 | 0.1626 |
| ER | 31 | 4 | 0.1884 |
| EK | 26 | 4 | 0.222 |
| C-Terminal Direction | | | |
| Pairs | Number of Genomes | n | $\phi_{XY}^{(n+)}$ |
| CC | 75 | 3 | 1.545 |
| LE | 58 | 2 | 0.2546 |
| ER | 58 | 3 | 0.2756 |
| LD | 36 | 2 | 0.1416 |
| LQ | 35 | 2 | 0.1357 |
| EK | 35 | 3 | 0.2849 |
| PC | 30 | 2 | 0.2229 |
| EM | 26 | 2 | 0.1542 |
| LN | 26 | 2 | 0.1231 |

### Table 2b: Under-represented First-Passage Times

| Significant First-Passage Times | | | |
|---|---|---|---|
| N-Terminal Direction | | | |
| Pairs | Number of Genomes | n | $\phi_{XY}^{(n-)}$ |
| GD | 29 | 2 | -0.1660 |
| C-Terminal Direction | | | |
| Pairs | Number of Genomes | n | $\phi_{XY}^{(n+)}$ |
| ED | 29 | 2 | -0.1725 |

**Supplementary Information**

**Supplementary Figures**

**Supplementary Figure 1: Predicted Amino Acid Frequency from Overall Coding Nucleotide Frequencies**

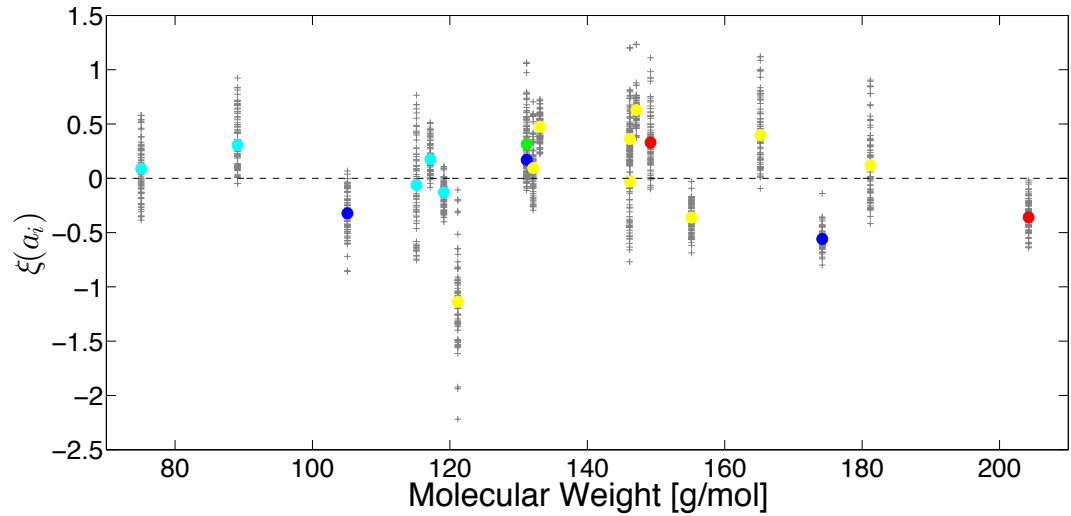

Logarithm of the ratio of the observed amino acid frequency to the predicted frequency based on the total frequency of nucleotides in all coding regions.

**Supplementary Figure 2: OGT as predicted by the relative abundance of CP**

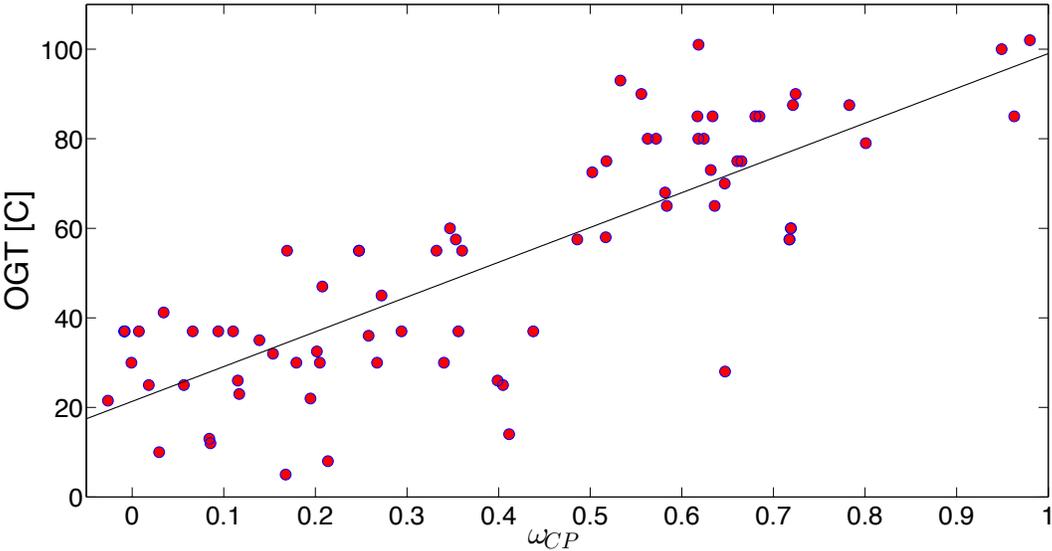

The optimal growth temperature of all 76 organisms versus the logarithm of the relative abundance of the dipeptide CP.